# Electronic structure and optical properties of Graphene Monoxide

Gui Yang, Yufeng Zhang, Xunwang Yan

(1) *College of Physics & Electrical Engineering, Anyang Normal University, Anyang Henan, 455000 China*

**Abstract:** The electronic and optical properties of graphene monoxide, a new type of semiconductor materials, are first theoretically studied based on density functional theory. Electronic calculations show that the band gap is 0.952 eV which indicate that graphene monoxide is a direct band gap semiconductor. The density of states of graphene monoxide and the partial density of states for C and O are given to understand the electronic structure. In addition, we calculate the optical properties of graphene monoxide including the complex dielectric function, absorption coefficient, the complex refractive index, loss-function, reflectivity and conductivity. These results provide a physical basis for the potential applications in optoelectronic devices.

*Email: kuiziyang@126.com

1. **Introduction**

Graphene is an exciting material for fundamental and applied solid state physics research due to its special electronic structure and the massless Dirac-fermion behavior[1-9]. Novel condensed matter effects arising from its unique two dimensional(2D) energy dispersion along with superior properties make it a promising material for the next generation of faster and smaller electronic devices. However, with 2D graphene being a zero-gap semiconductor, its use in an active electronic device such as a field effect transistor(FET) lacks an essential feature, namely, a band gap around the Fermi level. To solve this problem, various schemes have been presented to open a band gap in graphene[10-15]. One scheme proposed by Eduardo groups is that the band structure of bilayer graphene can be controlled externally by applying a gate bias, and extracted the value of the gap as a function of the electronic density by using a tight binding model[10]. Another suggestion is the creation of gaps through confinement, band gad engineering of graphene nanoribbons(GNRs) has been experimentally demonstrated, and GNRs-based field effect transistor(with a width of several tens of nanometers down to 2nm) have been characterized[12,13,15]. Chemical modification provides an effective way to manipulate the electronic properties of graphene-based materials. Doping graphene chemically with nitrogen have been studied both theoretically and experimentally[16-22].The interplay between nitrogen lone-pair electrons and the graphene $\pi$ system changed the electronic property and chemical reactivity. Recently, two stable, ordered N-doped graphene structures, $C_3N$ and $C_{12}N$ are theoretically revealed through cluster-expansion technique and particle-swarm optimization method[23]. Calculations show that the band gap of $C_3N$ and $C_{12}N$ are 0.96 ev and 0.98 eV, respectively which is a good band gap for many semiconductors.

Graphene-oxide(G-O) has recently emerged as a new carbon-based nanoscale material[24,25]. It contains a range of oxygen functional groups which renders it a good candidate for many applications, such as sensors and flexible transparent conductive electrodes[26]. However, previous studies have indicated that G-O was an insulator with poor electronic properties. Thus, control oxidation of G-O by various chemical and thermal reduction treatments plays an important role which provides tunability of the electronic and mechanical properties. Recently, a new material of graphene monoxide (GMO) has been experimentally developed which has quasi-hexagonal unit cell, an unusually high 1:1 C:O ratio[27]. The calculation based on density functional theory(DFT) showed that the direct band gap of GMO is ~0.9 eV. The semiconducting properties of GMO suggest that this new material might be useful for various electronic applications which may hold the key towards graphene based electronics. But the ground state properties and optical properties of this new type of semiconductor are still far from clear. In this paper, we first present a detailed theoretical study about the electronic and optical properties of GMO which is critical to the potential applications in electronic devices.

2. **Computational method**

Detailed description of the crystal structural data of GMO can be derived from Refs. [27]. Guided by the experiment, the lattice and electronic structure of GMO are calculated using DFT. The CASTEP code is first used to optimize the initial structure. The interactions between valence electrons and the ionic core were represented by ultra-soft pseudopotentials. The exchange-correation energy was calculated by the generalized gradient approximation (GGA) in terms of Perdew-Burke-Ernzerl (PBE). The valence electron configuration considered in this paper included O $2s^22p^4$ and C $2s^22p^2$. Lots of converging tests are performed, and with the cutoff energy of 380 eV and 10×10×1 Monkhorst-Pack k-points for integrals in the Brillouin zone. The structure is considered to be in equilibrium when the average force acting on ions was finally reduced to $5.0×10^{-7}$ eV/atom.

**3. Results and discussion**

3.1 Geometry structure and Muliken's population analysis

In order to ensure the accuracy of the calculation, we start our simulations with the case of graphene. The calculated lattice structure and band structure for graphene are shown in Fig.1. From Fig.1(b), one can see that the band gap of graphene is zero which shows that graphene is an zero-gap semiconductor. All of the optimize structure and calculated band structure for graphene are agreement with the theoretical results reported elsewhere, which confirm the accuracy and reliability of our calculation. Then, a detailed simulation is considered for GMO. We construct the crystal structure directed by the Ref[27] and optimize it by the CASTEP package. The Muliken's population analysis is firstly calculated to investigate the electronic structure of GMO as listed in table 1. The results about the bond parameters for GMO are marked in Fig.2(a). Compared with graphene, the bond length of C-C is changed from 1.42 Å to

1.661 Å and 1.918 Å, respectively. The primitive vectors increases from 2.46 Å to 3.09 Å, and the angle between them is $120^0$ versus the $124^0$. Thus, the structure of GMO no longer has the 3-fold symmetry of the graphene lattice. Consequently, the space group for graphene belongs to $D_{6h}$ symmetry while the GMO corresponds to $D_{2h}$ symmetry. In addition, it is evident from Table 1 that the charge of O and C in GMO is -0.45 and 0.45, which is related to electron gains and losses. For this reason, the total electron number for C 2s2p decreases from 4.0 to 3.55, and it increases to 6.45 for O 2s2p.

3.2 Band structure and density of states

The band structure for GMO from the GGA and local-density approximation(LDA) calculations are obtained, and the corresponding band gap is 0.952 eV and 0.811 eV, respectively. The former is more close to the calculated result ∼0.90 eV in Ref.[27] which is attributed to that LDA underestimates the band gap of semiconductors. Thus, we calculate the electronic structure by employing the GGA. Fig.2(b) is the band structure of GMO which shows that it is a direct band gap semiconductor. The valence band maximum and the conduction minimum are located at the G point. Fortunately, a band gap is opened in graphene-based material by the chemical functionalization which has potential applications in electronic devices. Fig.3 shows the total density of states(TDOS) for GMO and partial DOS plots of C and O, where line 1 represents the Fermi level. One can see that the peaks of TDOS on the left side of line 2 are mainly provided by O 2s and C 2s orbital. As the energy is greater than -14.2 eV corresponding to the right side of line 2, the curve in TDOS mostly consists of O 2p and C 2p orbital. Interestingly, the valance band near the Fermi level is mainly composed of O 2s2p, while the conduction band is attributed to C2s2p。

3.3 optical properties

The study of the optical functions helps to give a better understanding of electronic structure which can find potential applications in photoelectron devices and the semiconductor industry. The optical properties may be obtained from the complex dielectric function, $\varepsilon(\omega)= \varepsilon_1(\omega)+i\varepsilon_2(\omega)$. The imaginary part of the dielectric constant $\varepsilon_2(\omega)$ can be calculated from the momentum matrix elements between the occupied and the unoccupied electronic states. The real part $\varepsilon_1(\omega)$ is derived from the imaginary part $\varepsilon_2(\omega)$ by the Kramers-Kronig transform. Besides, other optical constants, such as absorption coefficient α, complex refractive index $n_c$= $n$-$ik$ with $n$, $k$ are the refractive index and extinction coefficient, loss-function β, reflectivity $R$ and conductivity σ are calculated by the flowing equations[28].

$$\varepsilon_2(\omega) = \frac{2e^2\pi}{\Omega\varepsilon_0} \sum_{k,v,c} |\psi_k^c|u.r|\psi_k^v| \delta(E_k^c - E_k^v - E) \tag{1}$$

$$\varepsilon_1(\omega) = 1 + \frac{2}{\pi} P \int_0^\infty \frac{\omega'\varepsilon_2(\omega')}{\omega'^2 - \omega^2} d\omega' \tag{2}$$

$$\alpha(\omega) = \sqrt{2}[\sqrt{\varepsilon_1^2(\omega)+\varepsilon_2^2(\omega)} - \varepsilon_1(\omega)]^{1/2} \tag{3}$$

$$n(\omega) = \frac{\sqrt{2}}{2}[\sqrt{\varepsilon_1^2(\omega)+\varepsilon_2^2(\omega)} + \varepsilon_1(\omega)]^{1/2} \tag{4}$$

$$k(\omega) = \frac{\sqrt{2}}{2}[\sqrt{\varepsilon_1^2(\omega)+\varepsilon_2^2(\omega)} - \varepsilon_1(\omega)]^{1/2} \tag{5}$$

$$\beta(\omega) = \text{Im}(\frac{-1}{\varepsilon(\omega)}) \tag{6}$$

$$R(\omega) = \left| \frac{\sqrt{\varepsilon(\omega)} - 1}{\sqrt{\varepsilon(\omega)} + 1} \right|^2 \tag{7}$$

$$\sigma(\omega) = \text{Re}(-i\frac{\omega}{4\pi}(\varepsilon(\omega) - 1)) \tag{8}$$

Where $u$ is the vector defining the polarization of the incident electric field, $\omega$ is the light frequency, $e$ is the electric charge and $\psi_k^c$ and $\psi_k^v$ are the conduction and valence band wave functions at $k$, respectively. The P in front of the integral means the principal value.

Fig.4 depict the dielectric function and the absorption coefficient of GMO for photon frequency up to 40 eV. The lines of solid and dot in Fig.4(a) represent the imaginary and real part of the dielectric function, respectively. It is observed that the static constant $\varepsilon_1(0)$ is close to 3, and the curve drops rapidly till $\omega=1.82$eV. In the range of 1.82-12.02 eV, $\varepsilon_1(\omega)$ rises slowly and then decreases sharply. Further increase the frequency, the value of $\varepsilon_1(\omega)$ increases slowly and up to a constant $\varepsilon_1(\omega)=0.97$ when $\omega>35$ eV. Two obvious peaks of the $\varepsilon_2(\omega)$ can be found which is always related to the electron exciation. The imaginary part of dielectric function $\varepsilon_2(\omega)$ vanishes with the frequency $\omega>24.9$eV. It is noteworthy that the value $\varepsilon_1(\omega)>0$ and $\varepsilon_2(\omega)=0$ means the region is a transparent area. Refer to the absorption parameter as shown in Fig.4(b), the optical absorption edge is about 0.952 eV that is corresponding to the band gap of GMO in our calculation.

The other optical constants including the complex refractive index $n$ and $k$, loss-function $\beta$, reflectivity $R$ and conductivity $\sigma$ are displayed in Fig.5. The static refractive index $n$ of GMO shown in Fig.5(a) has the value 1.7. In Fig.5(b), the electron energy function describes the energy loss of a fast electron transversing a material with the change of the frequency. Two apparent peaks of loss function locate at 1.80 and 11.93 eV, which are associated with the plasma frequency. Fig.5(c) shows the reflectivity spectra as a function of frequency. The spectra shows that the reflectivity $R$=0 when the frequency ω>24.9eV which proves once again that the existence of the transparent area. From to Fig.5(d), it depicts the change of the conductivity caused by the frequency which is the physical basis to the application in optical electronic devices. The curve is similar to the absorption coefficient in Fig.4(b) and the corresponding location for the peaks and troughs is the same. The minimum value is in 3.79eV while the maximum point occurs at 10.16 eV. As ω>24.9eV, its value is equal to zero.

4. Conclusion

In summary, the electronic structure and optical properties of GMO are studied based on the density functional theory. Directed by the experimental data of GMO, we use the CASTEP package to calculate the Muliken's population analysis and the band structure. The results show that the geometric structure changed significantly from graphene to GMO. As shown in Fig.1 and Fig.2, the magnitude of the primitive vectors increases from 2.46 Å in graphene to 3.09 Å and the angle between them is $124^0$ versus the $120^0$ of an ideal hexagonal lattice of graphene. In particular, the band structure indicates that GMO is semiconducting with the direct band gap of 0.952 eV. This new material has the potential for various electronic applications which may hold the key towards graphene based electronics. Further, the optical properties, e.g.

dielectric function, absorption coefficient, the complex refractive index, loss-function , reflectivity and conductivity are analyzed in detail.

**Acknowledgements**

This work was supported partly by the National Natural Science of China (Grant No. 11047108, 11147197, 11005003), partly by the Research Project of Basic and Cutting-edge Technology of Henan Province, (Grant No. 112300410183), and partly by the Education Department and Department of Henan Province (Grant No. 2011B140002 ).

Figure and table captions

Table 1 Population analysis results of graphene and GMO

| Material | Species | S(e) | P(e) | Total | Charge (e) | Bond | Length(Å) |
|---|---|---|---|---|---|---|---|
| Graphene | C | 1.05 | 2.95 | 4.0 | 0.0 | C1-C2 | 1.420 |
| GMO | O | 1.82 | 4.63 | 6.45 | -0.45 | C-O | 1.425 |
|  | C | 1.13 | 2.42 | 3.55 | 0.45 | C1-C2 | 1.661 |

| | | | | | | C3-C4 | 1.918 |
|---|---|---|---|---|---|---|---|

Fig.1. (a) The 3×3 graphene supercell with the calculated structural parameters. (b) The band structure for graphene with the band gap is zero.

Fig.2. Proposed GMO geometric structure and band structure. Carbon and Oxygen atoms are grey and red. (a1) Perspective view of 3×3 graphene monoxide supercell, and the top and side view for C-O. (a2) Top view of GMO and the detailed structural parameters. (b) The GGA calculated band structure for GMO with the band gap is 0.952 eV.

Fig.3. The total density of states of GMO and partial density of states for C and O.

Fig.4. The dielectric function and absorption coefficient of GMO.

Fig.5. The optical functions of GMO (a) complex refractive index (b) loss function (c) reflective (d) conductivity.

Fig.1

(a) Graphene

(b) [band structure plot: Energy (eV) vs G-K-M-G]

Fig.2

(a) Graphene Monoxide

(a1) Perspective view

Side View
Top view

(a2) Top view

(b) [band structure plot: Energy (eV) vs G-K-M-G]

Fig.3

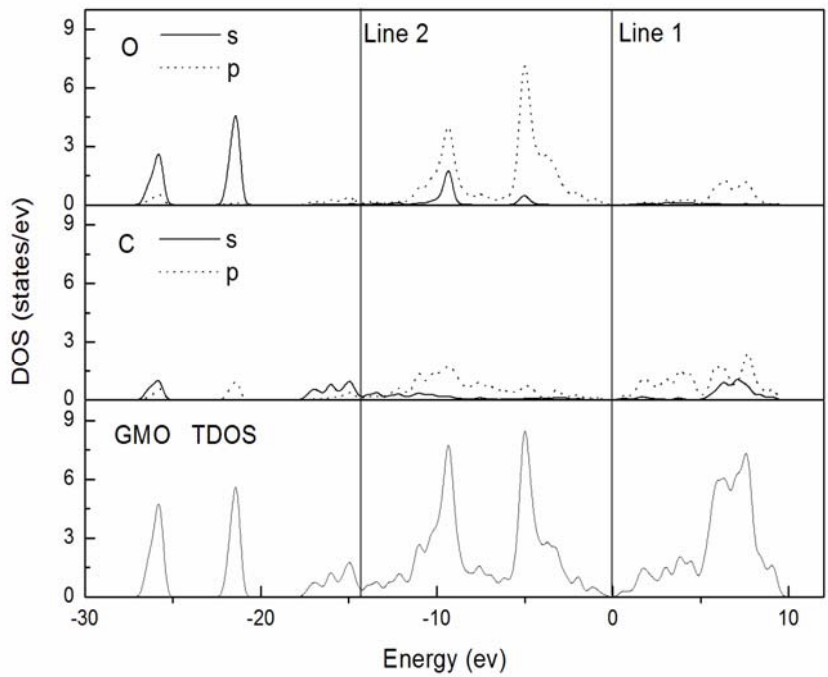

Fig.4

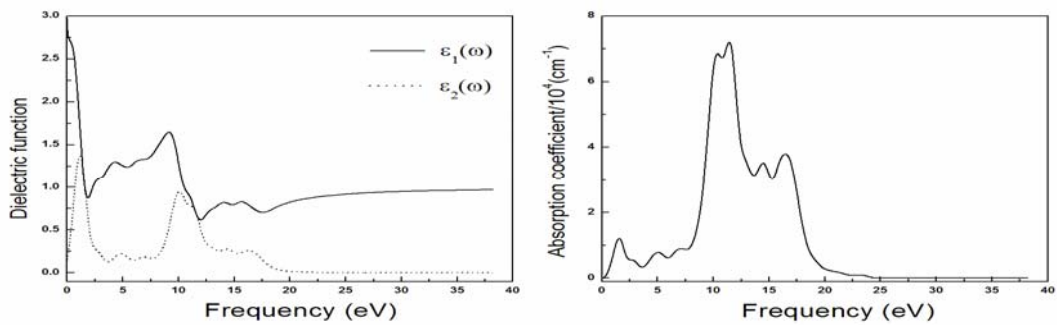

Fig.5

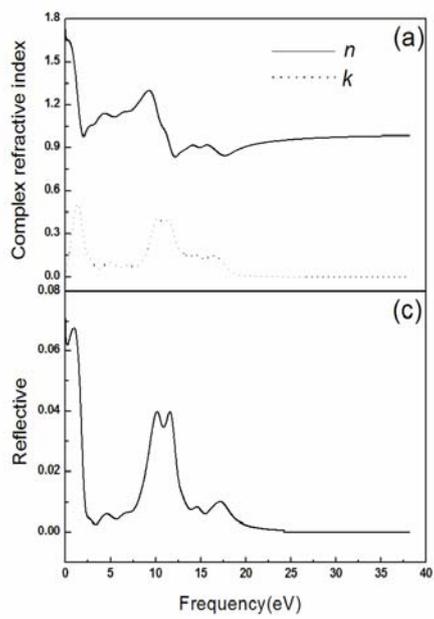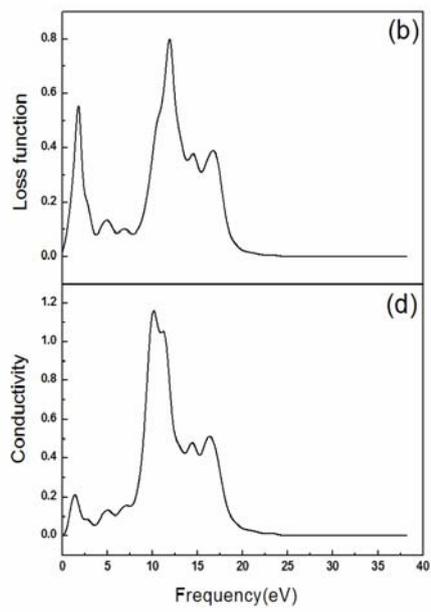